\documentclass[twocolumn,showpacs,preprintnumbers,amsmath,amssymb]{revtex4}

\usepackage{multirow}%
\usepackage{graphicx}
\usepackage{dcolumn}
\usepackage{bm}

\newcommand{\sfa}{SrFe$_2$As$_2$}

\newcommand{\fete}{Fe$_{1.087}$Te}

\begin{document}

\title{Strong coupling to magnetic fluctuations in the charge dynamics of Fe-based superconductors}

\author{J. N. Hancock$^{1}$}
\author{S. I. Mirzaei$^{1}$}
\author{J. Gillett$^{2}$}
\author{S. E. Sebastian$^{2}$}
\author{J. Teyssier$^{1}$}
\author{R. Viennois$^{1}$}
\author{E. Giannini$^{1}$}
\author{D. van der Marel$^{1}$}

\affiliation{$^{1}$
D\'epartement de Physique de la Mati\`ere Condens\'ee, Universit\'e de Gen\`eve, quai Ernest-Ansermet 24, CH 1211 Gen\`eve 4, Switzerland}
\affiliation{$^{2}$Cavendish Laboratory, Cambridge University, J. J. Thomson Avenue, Cambridge CB3 OHE, UK}

\date{\today}

\begin{abstract}
We present a comprehensive comparison of the infrared charge response of two systems, characteristic of classes of the 122 pnictide (\sfa) and 11 chalcogenide (\fete) Fe compounds with magnetically-ordered ground states. In the 122 system, the magnetic phase shows a decreased plasma frequency and scattering, and associated appearance of strong mid-infrared features. The 11 system, with a different magnetic ordering pattern, also shows decreased scattering, but an increase in the plasma frequency, while no clear mid-infrared features appear below the ordering temperature. We suggest how this marked contrast can be understood in terms of the diverse magnetic ordering patterns of the ground state, and conclude that while the high temperature phases of these systems are similar, the magnetic ordering strongly affects the charge dynamical response. In addition, we propose an optical absorption mechanism which appears to be consistent with information gained from several different experiments.
\end{abstract}

\pacs{PACS}

\maketitle

The iron pnictide or chalcogenide systems have often been compared to the cuprates based on proximity of magnetism to superconductivity. However, it is not clear whether these two classes of iron systems belong to the same universality class, given the notably different magnetic ordering pattern of the parent compound ground states.
To address these questions, we present an in-depth study of the charge dynamics of two systems with different magnetic ground states. Below $T_{ms}$=67 K, the ``11'' \fete\ undergoes a magnetostructural phase transition into a bi-diagonal magnetically-ordered ground state \cite{li09,bao09,chenelecfete09} (Figure 1a) with a large moment of $\sim$2$\mu_B$. In contrast, the ``122'' system \sfa\ ($T_{ms}$=190 K) orders in a smaller moment, vertical-stripe pattern \cite{huang08} (Fig 1b), common to the parent compounds of the type XFe$_2$As$_2$ and related compounds isostructural to LaPAsO. Both of these systems can be driven superconducting by chemical substitution \cite{hsu08,rotter08} and pressure \cite{han10,alireza09}. Formal valence counting gives the same nominal Fe valence for the two systems, while electronic structure \cite{subedi08,singhdu08,mazin08} and angle-resolved photoemission (ARPES) \cite{yi09,xia09} results suggest topological equivalence to the Fermi surfaces in the high temperature, paramagnetic phase.

Large samples of \fete\ and \sfa\ were grown using the Bridgeman and flux methods, respectively. Single crystalline platelets were then characterized by magnetometry, X-ray diffraction, and resistivity measurements. Electronic structure calculations were performed using the LMTart program \cite{sav95} and the experimentally determined atomic positions. Optical data were collected using a combination reflectivity and ellipsometry techniques. Figures 1e-f show the optical conductivity $\sigma_1(\omega)$ for \fete\ and \sfa\ above and below their magneto-structural transition temperatures. Strong infrared changes accompany the transition in \sfa, clear mid-infrared features appear at 550 and 1300 cm$^{-1}$ at the lowest temperatures, qualitatively consistent with previous studies \cite{hu08,wu09,chencaxis10}. Approaching $T_{ms}$ from below, a marked weakening and softening of these features is apparent. In contrast, \fete\ shows far weaker temperature dependence, and only a broad, flat background free of sharp peaks, similar to the high temperature phase of \sfa. 
LDA calculations for the high temperature phase predict more distinct features in $\sigma(\omega)$  (Fig. 1i) than the flat, smooth appearance of the experimental spectra. 
Interestingly, the conductance per layer, defined as $G(\omega)=\sigma_1(\omega)a_c$ where $a_c$ is the $c$-axis lattice parameter, is expected to be $(\pi/2)e^2/h$ \cite{gusynin06} for a Dirac cone spectrum as is observed in carbon \cite{kuz08}. This type of band structure is predicted \cite{morinari10,ran09} and observed \cite{harrison09,richards10} for 122 systems in the magnetic phase. Here, we observe a broad flat conductance of similar magnitude over a wide energy span at all temperatures, suggesting that short-range correlations are present in the paramagnetic phase as well.



\begin{figure}
\begin{center}
\includegraphics[width=3.4in]{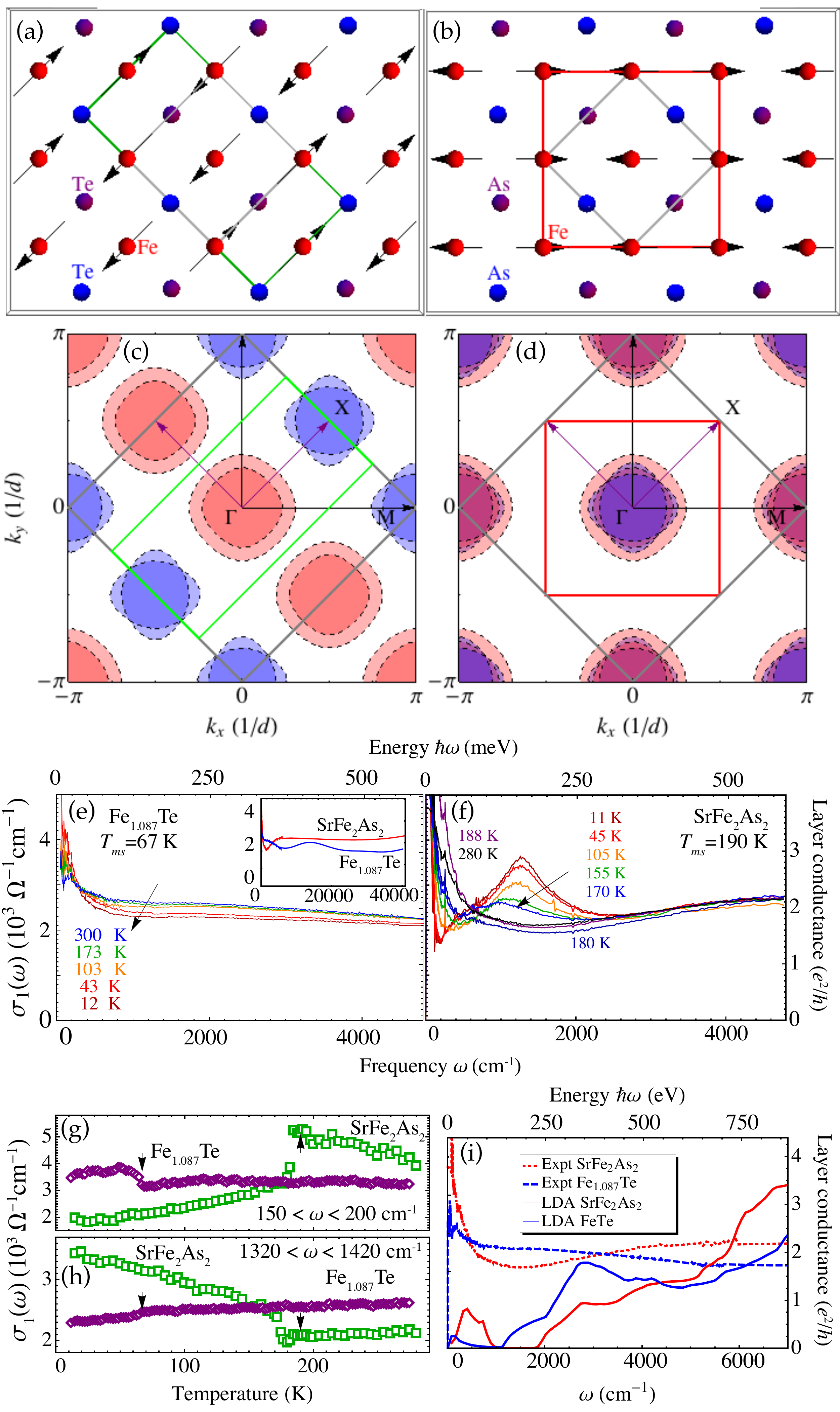}
\caption{Low-temperature magnetic structure of (a) \fete\ and (b) \sfa. (c,d) Fermi-surface folding expected from the lowered structural symmetry. Optical conductivity $\sigma_1(\omega,T)$ of (e) \fete\ and (f) \sfa\ crossing their magnetostructural phase transitions. Dashed lines in (e) inset marks the $\sigma_1$ corresponding to $G=(\pi/2)e^2/h$. Temperature dependence of $\sigma_1$ averaged over a narrow band of (g) far- and (h) mid-infrared frequencies. (i) Comparison of the high-temperature conductivity and the interband conductivity calculated from LDA \cite{sav95}.}
\label{ }
\end{center}
\end{figure}

\begin{figure}
\begin{center}
\includegraphics[width=3.5in]{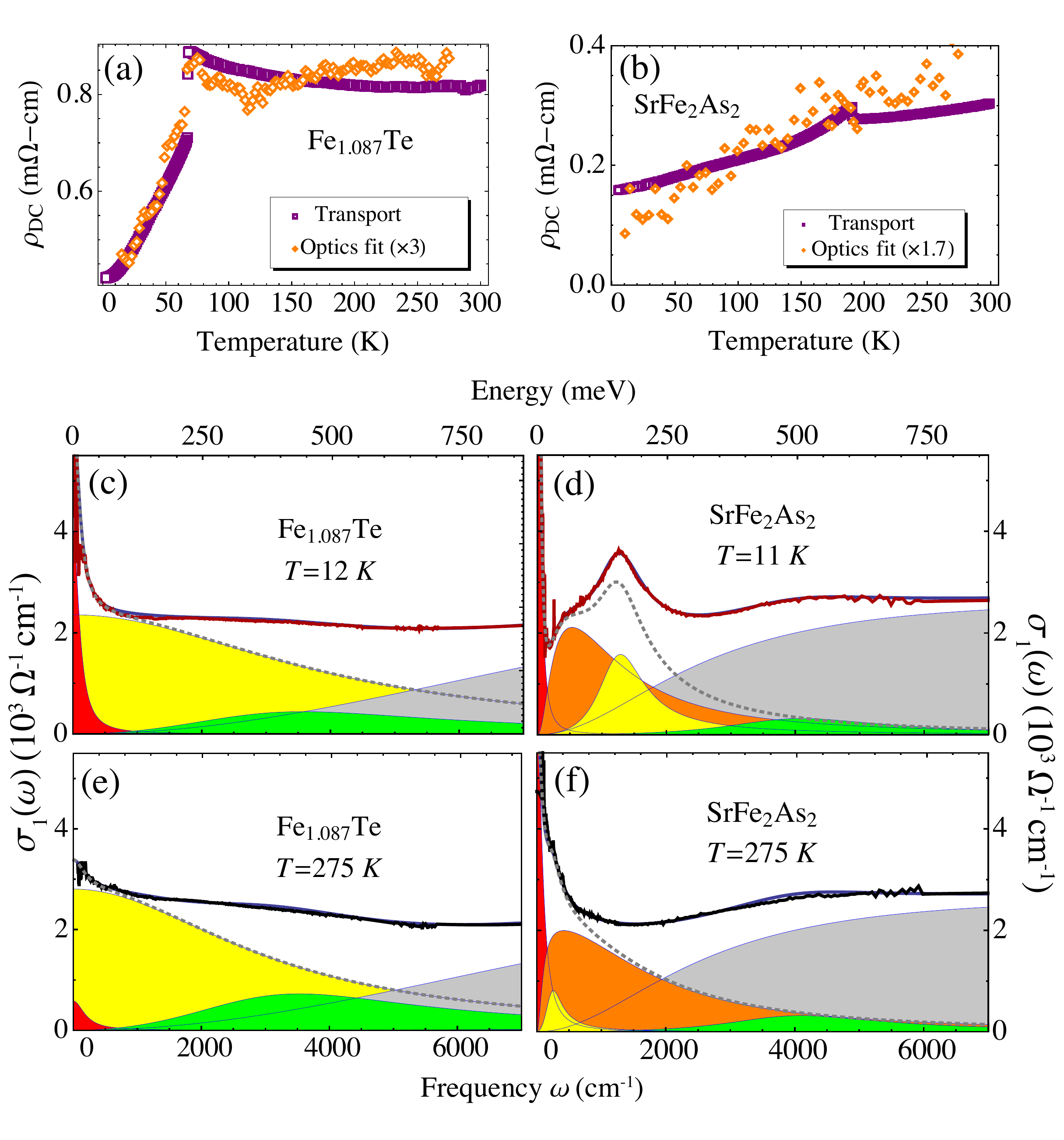}
\caption{DC resistivity measured by transport (squares) and optically (diamonds) for (a) \fete\ and (b) \sfa. Components of fits to the optical conductivity at high and low temperatures for \fete\ (c,e) and \sfa\ (d,f).}
\label{ }
\end{center}
\end{figure}

Figures 1g,h compare the temperature evolution of the conductivity in narrow bands around select frequencies. In each case, a kink at the transition temperature indicates that the optical spectra are influenced by the onset of the staggered magnetic order, implying a reconstruction of electronic bands in response to the development of the order parameter. While the direction of the effect is opposite for the two systems in both the mid- and far-infrared spectral regions, 
the low temperature resistivities (Figures 2a,b) both increase monotonically upon warming up to $T\sim T_{ms}$, above which a weaker temperature dependence is observed. Based on the resistivity data alone, it is unclear whether increased scattering or net depletion of carriers is responsible for the increasing resistivity. To quantify the dynamical extension of the transport behavior to finite frequency, we have fit the complex optical data to a set of Drude-Lorentz oscillators:
\begin{equation*}
\label{ }
\epsilon(\omega)=\epsilon_\infty+\sum_i\frac{\omega_{p,i}^2}{\omega^2-\Omega_i^2-i\gamma_i\omega}=1+\frac{4\pi i\sigma(\omega)}{\omega}
\end{equation*}
This type of model is convenient because with only a handful of parameters, one may construct a description of the data which satisfies the Kramers-Kronig relations and is therefore consistent with the principle of causality.

Figures 2c-f show a minimalist  application of this model to the low-temperature spectra. In \fete, a minimum of two Drude ($\Omega_i=0$) components were necessary to represent the low-temperature spectra, with broad, low-frequency interband conductivity contributions at 3500 cm$^{-1}$ and higher. The Drude contributions are clearly separable due to their very different widths $\gamma_i$, with $i \in\{1,2\}$. Only quantitative changes distinguish the model in the high and low temperature phases.

For \sfa, however, with $T<T_{ms}$, two components represent the double-peak structure in addition to a single Drude component. At $T>T_{ms}$, this double peak structure washes away, a phenomenon discussed in connection to the temperature-induced closing of one \cite{moon10,chencaxis10} or more \cite{hu08,wu09} spin density wave (SDW) gaps. In an unconstrained fit, the two peaks move to low energy as the temperature is raised and their sum resembles a single wide Drude component and a narrow component (dashed line), as is the case for \fete. Thus, at high temperatures, the optical spectra of \fete\ and \sfa\ are effectively only quantitatively different. As the high temperature phases of these systems are carried into a superconducting state by chemical substitution \cite{hsu08,rotter08} or application of pressure \cite{alireza09} or strain \cite{han10}, strong $(\pi,0)$ spin fluctuations are detected as precursors to superconductivity \cite{qiu09,christianson08}. In this way, it appears that the low energy physics of these systems is similar in their paramagnetic states, and the point of strongest departure in their low-energy behavior is along the magnetic phase boundary $T_{ms}$. One is then confronted with the challenge of understanding the differences in the magnetic ground states, and the influences that drive them. 

\begin{figure}
\begin{center}
\includegraphics[width=2.7in]{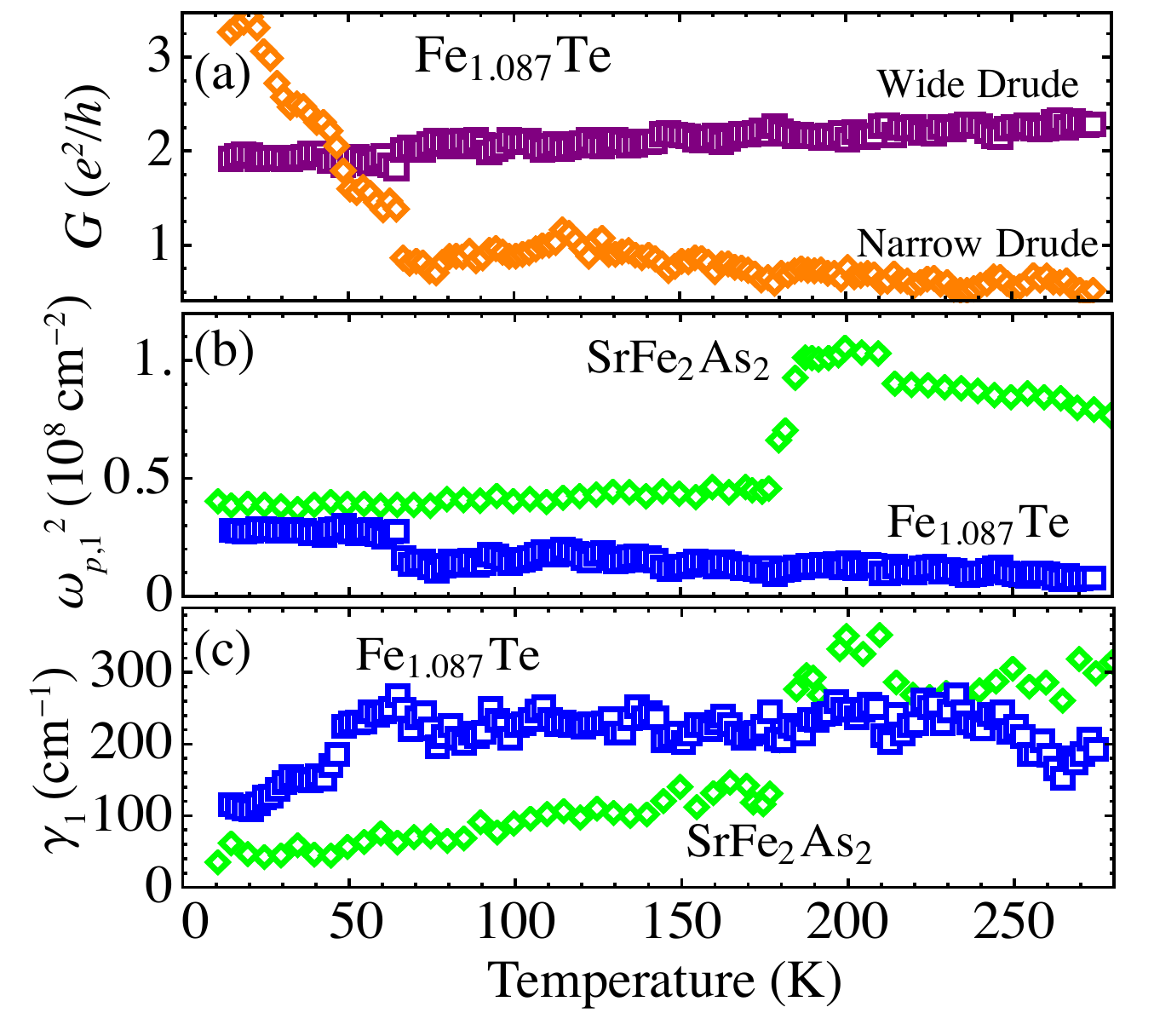}
\caption{(a) separate temperature evolution of the narrow and wide Drude contributions to the layer conductance of \fete. Comparison of the (b) strengths and (c) widths of the narrow Drude contributions for each of the two systems.}
\label{ }
\end{center}
\end{figure}

For \fete, Figure 3a shows separately the contributions of the two Drude components to the DC layer conductance $G(0)$. 
The wide Drude component has a nearly temperature independent value of 2$e^2/h$, and is unaffected by the magnetostructural transition. In our modeling scheme, nearly all of the temperature dependence in transport comes from the parallel conductance channel associated with the narrow Drude component, which grows significantly below $T_{ms}$. To better understand this behavior, Figure 3b compares the strength $\omega_{p,1}^2$ of the \textit{narrow} Drude component in each of the two systems. In each case, the value of the model $\omega_{p,1}^2$ has a step-like change at the magnetostructural transition, and is almost independent of temperature for $T<T_{ms}$. This reflects the fact that the structural transitions are 1st order \cite{li09,jesche08}, and the formation of electron and hole pockets occurs abruptly at $T_{ms}$, consistent with ARPES data on BaFe$_2$As$_2$ \cite{yi09} and Fe$_{1.06}$Te \cite{fengARPES10}. The direction of the step at the transition $T_{ms}$ is perhaps the clearest transport-related difference between these two systems: while the carrier weight in the 122 system decreases with the onset of magnetic order, the carrier weight of the 11 system increases. This result is robust against model details, and persists even if the low-$T$ two-peak carrier weight is completely transferred to carrier weight at high temperatures, as is suggested in the model fitting schemes of references \cite{hu08}, \cite{wu09}, and \cite{moon10}. This behavioral difference most likely arises from the diverse magnetic ordering pattern, which has profoundly different effects on the zone folding in reciprocal space (Figure 1c,d, and 4), and therefore also the topology of the low-temperature Fermi surfaces.

\begin{figure}
\begin{center}
\includegraphics[width=3.6in]{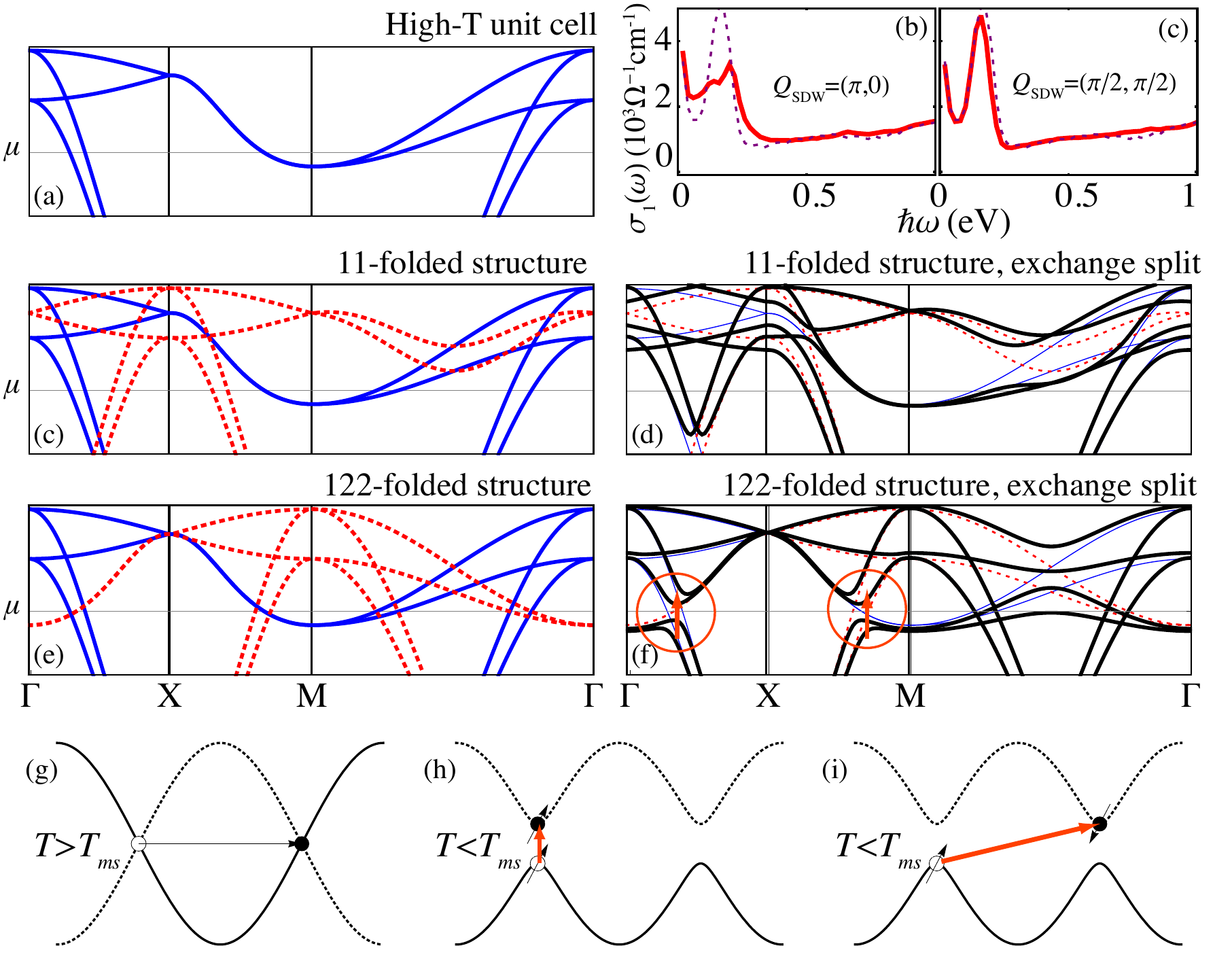}
\caption{Toy model band structure \cite{raghu09} for (a) high temperature unit cell, (c) folded at $Q_{AF}$=$(\pi/2,\pi/2)$ and (e) $Q_{AF}$=$(\pi,0)$. (d) and (f) respectively correspond to (c) and (e) with a small mean field mixing included. Allowed optical transitions are indicated by arrows in (f). In (b) and (c), show results of a 5-band tight-binding model calculation in the high temperature (dotted) and magnetic (solid) phases with ($\pi$,0) and ($\pi/2$,$\pi/2$) folding, respectively. (g) Optical process resulting in a damped fluctuation. (h) Direct  and (i) magnon-assisted indirect SDW gap crossing excitation.}
\label{ }
\end{center}
\end{figure}

We now relate the qualitative behavior of the higher-frequency optical spectra to these magnetically-induced topological changes. Figure 4a shows a toy model \cite{raghu09} band structure for the high temperature phase. Figures 4c-f show the effect of translation and mixing of these bands by the antiferromagnetic wave vectors for the 122 ($Q_{SDW}$=$(\pi,0)$) and the 11 ($Q_{SDW}$=$(\pi/2,\pi/2)$) magnetic structure. The 122 folding clearly results in more crossed bands near the Fermi level, due to the direct overlap of hole and electron pockets. At particular degeneracy points along the high-temperature Fermi surface, even small mixing and splitting can open momentum-conserving optical transition pathways, which appear near 550 cm$^{-1}$ in \sfa\ at low $T$. If only vertical transitions in the bandstructure (i.e. those transition whereby precisely one electron and one hole with overall momentum zero) are considered, the optical spectra should be only weakly influenced by the SDW order, as shown in Fig. 4b,c based on a 5-band tight binding model \cite{graser09} simulation with a mean-field exchange potential of 200 meV. Thus, while the different ordering patterns, low temperature moments, and transition temperatures of these two systems suggest a potentially different strength of the effective exchange potential, the 11 folding has an intrinsically smaller effect on the optical response than the 122 folding, and the salient contrast in optical response can be understood in this way.

We already related the smooth appearance of $\sigma(\omega)$ to strong coupling of electron and hole excited states to spin-fluctuations. Since the spin-fluctuation spectrum of an itinerant (anti-)ferromagnet undergoes drastic changes through the magnetic phase transition, this change should then become visible in the optical spectra. Since in the 122 system $T_{ms}$ is high, and the magnetism is typical for a metallic and itinerant material, the largest effect could be expected in this system as compared to \fete, as is indeed the case. The 1300 cm$^{-1}$ feature in the 122 systems have a particularly strong, order-parameter-like temperature dependence below $T_{ms}$ (Fig. 1h). Similar trends have been observed in XFe$_2$As$_2$ with X=(Ba, Sr\cite{hu08}, Eu\cite{moon10,wu09}). In Ba-122, a low-frequency peak is observed for polarization along and perpendicular to the $c$ axis, but the stronger, high frequency feature is observed only for polarization $\perp c$ \cite{chencaxis10}. Furthermore, across systems with different $T_{ms}$, the low frequency gap is of appropriate magnitude to consider a mean field BCS-type SDW gap crossing (pair-breaking) excitation. As pointed out before \cite{wu09,hu08,moon10}, the high energy feature, which appears at 1300 cm$^{-1}$ in \sfa, is highly anomalous.

Based on these considerations, we suggest a magnetic origin for the anomalous feature. Figure 4g shows a simplified schematic of the high-temperature band structure. Near the magnetic phase boundary, spin fluctuations provide a dissipation channel which involves an electron traversing a Fermi pocket in a Landau damping process. Below $T_{ms}$, hybridization and splitting of bands opens an SDW gap excitation pathway, giving rise to the 550 cm$^{-1}$ excitation (Fig. 4h). We assign the second excitation to a process shown schematically in Figure 4i. Here, a down-spin electron in an occupied state at momentum $-k_F$ undergoes a similar gap-crossing excitation to a state at +$k_F$, while flipping its spin. This is normally disallowed due to the photon momentum and spin selection rules, but can become allowed through the simultaneous emission of a $S=1$ magnon of momentum $-2k_F$, so that the overall momentum and spin of the process is conserved. Zhao \textit{et al} \cite{zhaoprl08} have shown that sharp quasi-2D spin waves with velocity $v_s\simeq$0.28 eV$\AA$ exist in the low temperature phase of \sfa. Yi \textit{et al} \cite{yi09} have detected strong SDW gapping along a hole pocket at $\Gamma$ and electron pocket at $M$, with gapped section connected by wavevectors $k_c$ traversing between 0.2-0.5 $\pi/a$ in the 2D-projected Brillouin zone. 
Thus, while the average SDW gap excitation appears near 550 cm$^{-1}$, and is roughly consistent with the BCS expectation $2\Delta\sim 3.5 k_BT_{ms}\sim$462 cm$^{-1}$, we expect the magnon-assisted pair breaking absorption to occur near $k_{c}v_s\sim$ 360-900 cm$^{-1}$ higher, which wholly encompasses the observed splitting between the high and low energy features in \sfa. In this scenario, the low temperature characteristics of the high energy peak in \sfa\ are expected to evolve with temperature as a direct consequence of broadened spin excitations as the temperature is raised, as observed. For $T>T_{ms}$, one expects an overdamped paramagnon response, thus explaining the difficulty in tracking this feature into the high-temperature phase. 

In summary, we have observed the influence of spin fluctuations on charge dynamics of two systems closely related to Fe-based superconductors. The spin-charge coupling is most evident at the magnetostructural transitions in these compounds, where transport and charge excitations are sensitive to drastic changes in the spin susceptibility. These observations are robust and likely extend to the superconducting portions of the global phase diagram for Fe-based supercodnuctors.

We acknowledge valuable conversations with A. B. Kuzmenko, I. I. Mazin and D. Parker. This work was supported by the Swiss National Science Foundation through the NCCR ÔÔMaterials with Novel Electronic PropertiesÕÕ (MaNEP). 

\bibliography{feHTSC}
\end{document}